\documentclass[journal]{IEEEtran}
\usepackage{amsmath}
\usepackage{graphicx}
\usepackage{mathtools}
\usepackage{epsfig}
\usepackage{epstopdf}
\usepackage{multirow}
\usepackage{array}
\usepackage{color}
\usepackage{multicol}
\usepackage{subfigure}
\usepackage{algorithmic}
\usepackage{booktabs}
\begin{document}

\title{Measuring Fatigue of Soldiers in\\ Wireless Body Area Sensor Networks}

\author{N. Javaid$^{1}$, S. Faisal$^{1}$, Z. A. Khan$^{2}$, D. Nayab$^{3}$, M. Zahid$^{4}$\\\vspace{0.4cm}
%    \IEEEauthorblockA{ \{nadeem.javaid,djouani@univ-paris12.fr\}}\\
        $^{1}$COMSATS Institute of IT, Islamabad, Pakistan.\\
        $^{2}$Dalhaousie University, Halifax, Canada.\\
        $^{3}$NWFP UET, Peshawar, Pakistan.\\
        $^{4}$Sarieddine Trading Est, Abu Dhabi, U.A.E.\\

     }

\maketitle

\begin{abstract}
Wireless Body Area Sensor Networks (WBASNs) consist of on-body or in-body sensors placed on human body for health monitoring. Energy conservation of these sensors,
while guaranteeing a required level of performance, is a challenging task. Energy efficient routing schemes are designed for the longevity of network lifetime.
In this paper, we propose a routing protocol for measuring fatigue of a soldier. Three sensors are attached to soldier's body that monitor specific parameters.
Our proposed protocol is an event driven protocol and takes three scenarios for measuring the fatigue of a soldier. We evaluate our proposed work in terms of
network lifetime, throughput, remaining energy of sensors and fatigue of a soldier.
\end{abstract}

\section{Introduction}
Wireless Sensor Networks (WSNs) seek the attention of researchers due to their effectiveness in multiple applications. A particular subclass of WSNs known as
WBASN consists of multiple sensors attached with the human body to provide us real time feedback like temperature, heartbeat, pulse rate and ECG monitoring.
Through WBASN a patient is monitored, and in case of critical situation an immediate action is made possible. Sensors collect data from the body of a patient
and send it to physician. The primary application of WBASN is continuous healthcare monitoring.

In WBASNs, monitoring of environment is a challenging task due to limited number and sensitive placement of sensors. Misplacement of sensors cause degradation
in the quality of captured data. So, placement of tiny and light powered sensors is an important factor. To prolong the lifetime of sensors, route selection is of
key importance. Thus, Authors in~\cite{1},~\cite{2} and~\cite{3} proposed energy efficient routing protocols. A common method for maximizing the sensors lifetime is
minimization of communication between sensors. For energy consumption minimization~\cite{4}, special attention must be given to enhance the communication system.

Fatigue, physical or mental, is a subjective feeling of tiredness. Sometimes, it is correlated with lethargy. Physical fatigue is the inability of muscles to
maintain optimal physical performance. Medically, fatigue is considered as a symptom rather than a sign because it is reported by a patient in a subjective manner.
Normally, fatigue is caused by loaded work, depression, boredom, mental stress, lack of sleep, etc~\cite{5}. Different routing protocols have been proposed for
 various data demands in WBASNs. If sensors sense and gather data constantly and transmit it periodically then this type of communication is called clock driven
 communication. In event driven communication, transmission is triggered by a particular event. Query driven communication deals with the transmission
occurrence in response to a query.

Various techniques are proposed for improving the efficiency of direct communication. In this paper, we present a new routing scheme for measuring the fatigue
 of a soldier. Our routing protocol takes three scenarios into account: (1) walking (2) slow running and (3) fast running. We use an event driven approach i.e.
 transmission is triggered by a particular event.

\section{Related work and Motivation}
%Many protocols have been proposed for WSNs. However, these protocols are not well suited for WBASNs.\\

Authors in~\cite{6} proposed Energy-Balanced Rate Assignment and Routing protocol (EBRAR) for body area networks. In EBRAR, routes are selected on the basis of
residual energy of nodes. Moreover, intelligent data transmission and uniform distribution facilitate network lifetime enhancement.

In~\cite{7}, authors proposed an opportunistic routing scheme for WBASNs. Quality of link between nodes in WBASN is effected by the body movement. Therefore,
proposed opportunistic routing protocol maximizes the lifetime of sensors during body movements.

N. Javaid \textit{et al.}~\cite{8} proposed a routing protocol for heterogeneous WBASNs. Data is transmitted directly for on-demand data and multi-hop
communication is used for normal data delivery in the proposed protocol.

Authors in~\cite{9} introduced $3$ types of nodes with different energies, i.e. normal, advanced and super nodes. Energy of super nodes is more than normal
nodes and advanced nodes. CHs are selected on the basis of their energies.

In Q-LEACH~\cite{10}, the network is divided into four quadrants. Each quadrant possesses specific number of nodes, and is further divided into sectors.
Each sector selects its own CH and hence load on CH reduces.

HEER~\cite{11} is a cluster based reactive routing protocol which uses residual energy of nodes and average energy of the network for the selection of CH.
Furthermore, introduction of hard and soft threshold helps to conserve energy of the sensors.

A. Ahmad \textit{et al.} proposed DDR (Density controlled Divide-and-Rule) scheme for energy efficient routing in wireless sensor networks)~\cite{12}. In DDR,
static clustering technique is used. Nodes are distributed uniformly in the network and randomly in the clusters. In DDR, network is divided into segments.
Each segment is designed such that distance between CH and nodes, and between CH and base station is reduced. After every round new CH is selected in each
segment with the ability of fixed number of CH in each round. CH is selected such that its distance from central reference point is minimum. In every segment
mutihop communication strategy is adopted.

Authors in~\cite{13} present a system in which multiple sensors are attached with the body to monitor various body parameters such as, heart activities.
It is a radio based wireless network technology.

Authors in [14] conducted a comprehensive survey on different architectures used in WBASNs for ubiquitous healthcare monitoring.
Various standards and devices used in these architectures, and finally the influence of path loss in WBANs are also provided.

According to our knowledge, existing routing protocols in WBASNs  are mainly concerned with energy efficiency of nodes. Thus, we propose a routing
protocol to consider specific real time mobility scenarios like walking, slow running and fast running with an additional capability to measure fatigue.

\section{Proposed Protocol}
 For the improvement of security, we must know the physical status of soldiers. An important factor in physical status of soldier is fatigue. If we know the state
  of fatigue in soldiers then we can send backup to them. In this proposed protocol, we measure the fatigue of a soldier through WBASN. Sensors collect several
  body parameters of a soldier, send this data to Base Station (BS) placed on body and BS sends it back to headquarter. The main problem in WBASNs is the limited
  energy of sensors. So, an efficient routing protocol is needed for maximizing the lifetime of sensors. In this paper, we propose a protocol for measuring the
  fatigue of a soldier and also a routing protocol for maximizing the lifetime of sensors.
\subsection{Placement of sensors}
We place three sensors on the body of a soldier, each of which has a specific attribute to sense. Sensors that are placed on the body are:

\begin{itemize}
\renewcommand{\labelitemi}{$\rightarrow$}
\item Temperature sensor
\item Blood glucose level sensor
\item Heartbeat sensor
\item BS
\end{itemize}

\subsubsection{Temperature sensor}
This sensor is placed on the fingertip of soldier to measure the temperature.
\subsubsection{Blood Glucose level sensor}
The purpose of this sensor is to check glucose level in the blood of soldier. It is positioned on fingertip. As glucose level can only be checked by taking
blood samples, so we assume that sensor takes the blood samples periodically and checks glucose level in the blood.
\subsubsection{Heartbeat sensor}
This sensor is placed on heart of soldier to measure heartbeat.
\subsubsection{BS}
BS is placed on the wrist of soldier. We assume that BS do not have battery problem and is enriched with energy. All the above three sensors (temperature, blood
 glucose level and heartbeat) send data to BS. Fig. 1 shows the placement of nodes on soldier.
\begin{figure} [!ht]
\centering
\includegraphics[height=6cm,width=3.5cm]{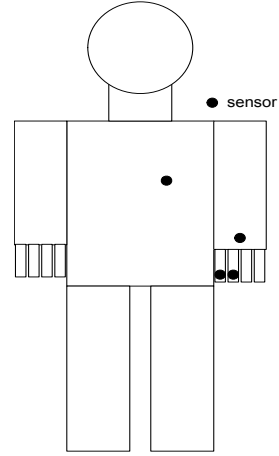}
\caption{Placement of nodes on soldier's body}
\end{figure}

\subsection{Scenarios}
Following are the scenarios for measuring the fatigue of soldier:

\begin{itemize}
\renewcommand{\labelitemi}{$\rightarrow$}
\item Soldier is walking
\item Soldier is running slowly
\item Soldier is running very fast
\end{itemize}

\subsubsection{Walking}
When soldier is walking, temperature of the body increases slowly. We assume that initially body temperature of soldier is normal. When soldier starts to walk,
temperature sensor starts to sense the body temperature. Initially, transmitter of sensor does not transmit data to BS because the body temperature of soldier is
slowly increasing and is not in danger. Sensor remains in sleep mode for a specific period of time. Whenever body temperature crosses threshold limit, transmitter
sends data to BS.

Heartbeat sensor checks the heart rate of soldier. At the beginning, heartbeat of soldier is normal. When soldier starts to walk, heartbeat slowly increases.
Sensor remains in sleep mode for normal heartbeat. Heartbeat sensor is triggered whenever palpitations cross a predefined threshold level.

Blood glucose level sensor measures glucose level in the blood of soldier. Blood glucose level of a person decreases while walking. We assume that sensor takes
blood samples periodically and compare the sensed value with threshold that is predefined. If the sensed value decreases from the threshold then sensor sends data
to BS which is an alarm that glucose level in blood of soldier is less.

During walk, temperature and heartbeat of soldier increases and glucose level in blood decreases with time. This means that energy of soldier is being consumed.
Sensors sense the values and send it to the BS that is placed on wrist of the soldier. BS checks received data and also the current energy of soldier. If state of
 fatigue is achieved then BS sends data to head quarter. For fatigue state we use Harris Benedict Formula. i.e.
\begin{equation}
    A=(13.75 \times weight( kg))
\end{equation}
\begin{equation}
    B=(5.003 \times height (cm))
\end{equation}
\begin{equation}
    C=(6.775 \times age (years))
\end{equation}
\begin{equation}
    BMR=66.5 + A + B - C
\end{equation}

%
%\begin{equation}
%BMR=66.5 + (13.75 \times weight( kg)) + (5.003 \times height (cm)) - (6.775 \times age (years))
%\end{equation}

Basal Metabolic Rate (BMR) is the amount of energy required to maintain normal metabolic rate of the body. This is the energy required for the functioning of vital
body parts like heart, lungs, kidneys, liver, nervous system and skin. We take this amount of energy as minimum energy of soldier and if a soldier's condition meets
 this level then this is called as `state of fatigue'.
\subsubsection{Slow Running}
If soldier is running slowly then sensors sense heartbeat, temperature and glucose level in blood. However, in this scenario the heartbeat, temperature and glucose
level reach the threshold earlier than walking.
\subsubsection{Fast Running}
In this scenario soldier is running at very high speed. So, in this case the fatigue state comes earlier than walking and slow running.

 %\subsubsection{Threshold for sensors}
\section{Radio Model for Transmission}
Here we discuss how much energy of sensors is consumed during transmission. In this paper, we use the energy consumption model used in~\cite{15}, and is summarized
 in the following equations.

\begin{equation}
   E_{tx}=E_{TXelec}\cdot k + E_{amp}\cdot(n)\cdot k \cdot d ^n
\end{equation}
\begin{equation}
   E_{rx}=E_{RXelec}\cdot k
\end{equation}

Where, $E_{tx}$ is the transmission energy, reception energy $E_{rx}$, $E_{TXelec}$ denotes the energy dissipated by radio to run the circuitry for the
transmitter, $E_{RXelec}$ represents the energy dissipated by radio to run the circuitry of the receiver, $E_{amp}$ is the energy for transmit amplifier,
$n$ is the path loss exponent. Its value is $3.38$ for line of sight communication and for non line of sight communication its value is $5.9$. $k$ shows the
 number of transmitted bits. Heartbeat sensor transmits $240$ bits, while blood glucose level sensor and temperature sensor transmit $2400$ bits data~\cite{15}.

 \begin{table}[!ht]
\begin{center}
\caption{Radio model parameters}
  \begin{tabular}{|c|c|}\hline
   \textbf{Parameters}                                           & \textbf{Value}            \\ \hline \hline
    \textbf{Initial energy $E_{o}$ }                             &  0.3 J                     \\ \hline
   %\textbf{Initial energy of advance nodes}                     &  $E_{o}$(1+$\alpha$)        \\ \hline
  %\textbf{Energy for data aggregation $E_{DA}$}                 & $5 nJ/bit/signal$            \\ \hline
  \textbf{Transmitting and receiving energy $E_{TXelect}$}       &  $ 16.7nJ/bit       $         \\ \hline
      \textbf{Transmitting and receiving energy $E_{RXelect}$}   &  $ 36.1nJ/bit       $          \\ \hline
      \textbf{Amplification energy $E_{amp}$}                    &  $1.97 nJ/bit/m^2$              \\ \hline
      %\textbf{Amplification energy for long distance $E_{amp}$} &  $0.013 pJ/bit/m^4$              \\ \hline
      %\textbf{probability  $p_{opt}$}                           & 0.1                              \\ \hline
\end{tabular}
\end{center}
\end{table}

Fig. 2 shows the transmission of data from nodes to server. Firstly, nodes sense data then they transmit it to the BS on the wrist of soldier. BS then transmit
 received data to server through mobile station.

\begin{figure*} [t]
\centering
\includegraphics[height=7cm,width=8cm]{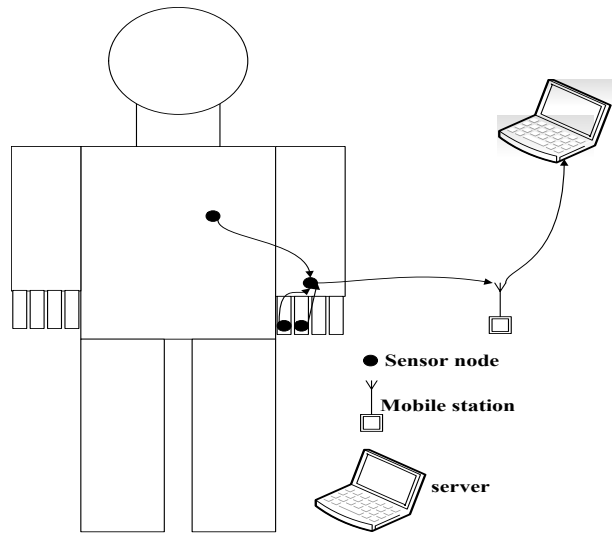}
\caption{Transmission of Data}
\end{figure*}

\section{Simulation and Results}
We use MATLAB~\cite{16} for simulation purpose. Radio model parameters used in simulations are shown in table 1. For simulations purpose, nodes are distributed
as shown in Fig. 1. BS is placed on the wrist of soldier. All the three nodes send data to BS directly if threshold value is satisfied. We take $5$
simulations to find average for each scenario and plot average results with $90\%$ confidence interval. In simulations we assume that walking speed of soldier
is $3.0$ miles per hour, slow running scenario speed is $5.0$ miles per hour and fast running the speed of soldier is $7.0$ miles per hour. Our goals in
conducting simulations are,
\begin{itemize}
\renewcommand{\labelitemi}{$\rightarrow$}
\item Measuring the fatigue of the soldier while walking, running slowly and running very fast.
\item Extension of node lifetime.
\end{itemize}

 Fig. 3 shows graph of alive nodes verses rounds. Round means network operation time in which nodes send data to BS. We assume that duration of a round is one second.
 In Fig. 3 , we can see that in walking scenario the lifetime of nodes is more than slow running and fast running scenarios. If soldier is walking, his/her activity
  is less and nodes do not send data more often. From activity we mean heartbeat, body temperature and glucose level of blood. Fig. 3 shows that sensor
  lifetime of walking scenario is more than slow running and fast running scenarios because, when a soldier is running the heartbeat and temperature increases and
   glucose level decreases quickly. So in fast running scenario, nodes send data to BS after very short interval of time which decreases the lifetime of
   nodes. Similarly in slow running scenario nodes send data to BS when nodes detect change in heartbeat, temperature and glucose level in blood which is
   comparatively less then fast running scenario. It results in better network lifetime for slow running scenario as compared with fast running scenario.
 \begin{figure} [!ht]
 \centering
 \includegraphics[height=7cm,width=9cm]{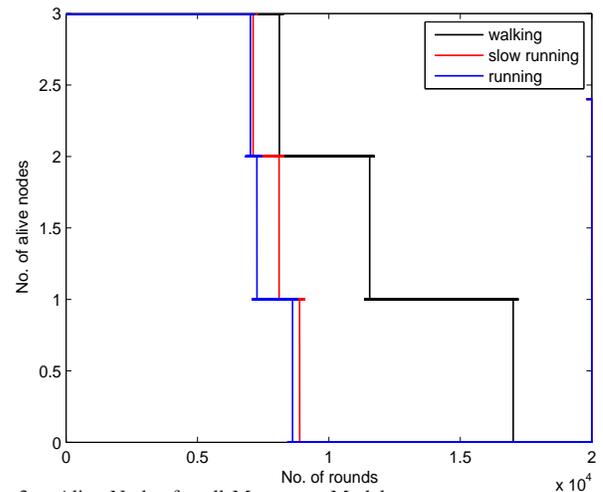}
 \vspace{-1cm}
 \caption{Alive Nodes for all Movement Models}
  \end{figure}

Throughput of nodes is shown in Fig. 4. We assume that the number of packets sent to BS are successfully received. In Fig. 4, we can see that when soldier is
walking, throughput is less as
 compared with slow running and fast running scenarios. Because, throughput is related with the heartbeat of soldier and heartbeat has a direct relation with
 the body temperature. So, in first scenario when a soldier is walking, his/her heartbeat is normal and throughput is less. Now, when the soldier is running fast,
 his/her temperature raises because of increased heartbeat and number of packets sent to BS increases.

 \begin{figure} [!ht]
 \centering
 \includegraphics[height=7cm,width=9cm]{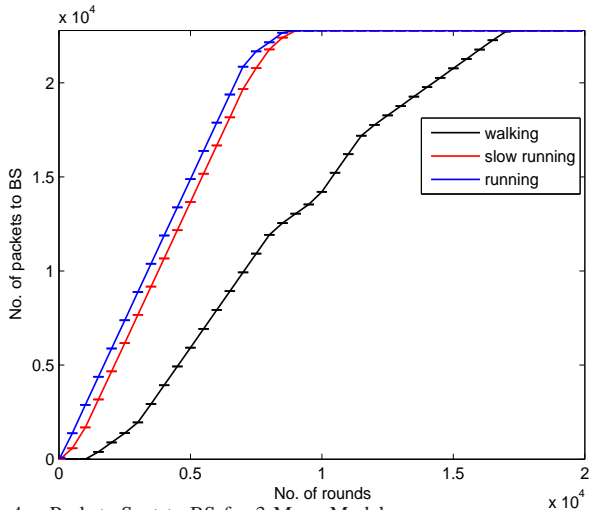}
 \vspace{-1cm}
 \caption{Packets Sent to BS for 3 Move Models}
  \end{figure}

In measuring the fatigue of soldier we set a threshold by using the Harris Benedict Formula. The threshold 1500 (joules) is set in our simulations. Fig. 5
shows the fatigue of soldier. We see that when soldier is walking, his state of fatigue comes later compared with slow and fast running scenarios because the
 activity of soldier's body is less and the energy is not consumed at faster rate. In fast running scenario the heartbeat and temperature increases at a faster
 rate which decreases the glucose level continuously. Therefore, soldier reaches his fatigue level much earlier.

 \begin{figure} [!ht]
 \centering
 \includegraphics[height=7cm,width=9cm]{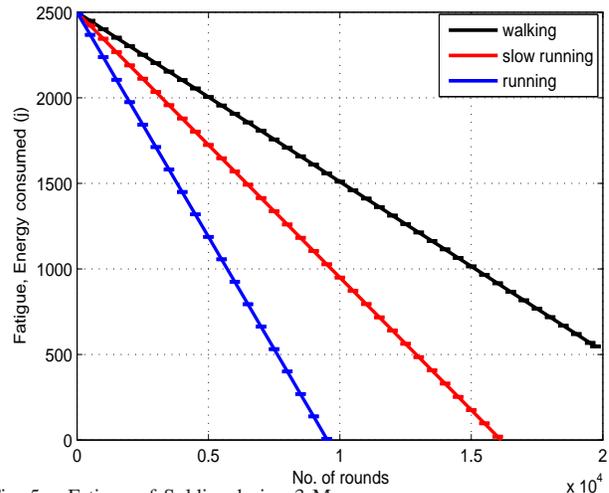}
 \vspace{-1cm}
 \caption{Fatigue of Soldier during 3 Moves}
  \end{figure}

  Fig. 6 shows the remaining energy of nodes in the network. In network every node has $0.3$ joules energy. So total energy of nodes in the network is $0.9$ joules.
   In fast running scenario, the activities of soldier are much greater than the walking scenario, and sensors are busy in getting the data from soldier's body and
   drain off their energy more quickly than the walking scenario. It can be seen that in slow running scenario the energy consumption is a bit less than fast
   running scenario.
  \begin{figure} [!ht]
 \centering
 \includegraphics[height=7cm,width=9cm]{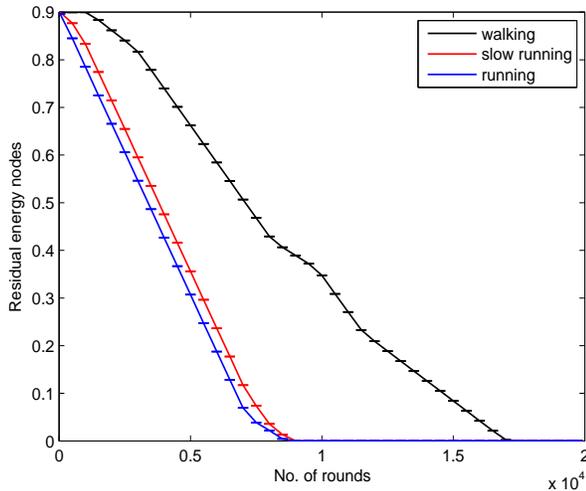}
 \caption{Residual Energy of Sensors for 3 types of motions}
  \end{figure}

 Table 2 shows analytical comparison of the nodes in three scenarios. From table we can see lifetime of the nodes in walking scenario is more than the slow and
  fast running scenarios.
  \begin{table}[!ht]
\begin{center}
\caption{Comparison of lifetime and throughput}
 \begin{tabular}{|p{1.5cm}|p{2cm}|p{2cm}|p{2cm}|}\hline
   \textbf{Protocol}           & \textbf{First node dead (round)} & \textbf{Last node dead (round)}  & \textbf{Throughput} \\ \hline \hline
    \textbf{Walking }          &  8116                         & 17010                            & $2.22\times10^4$\\ \hline
     \textbf{Slow running}     &  7114                         & 8854                             & $2.27\times10^4$\\ \hline
      \textbf{Fast running}    &  7015                         & 8617                             &$2.28\times10^4$\\ \hline
\end{tabular}
\end{center}
\end{table}

In simulations, we assume total energy of soldier to be 2500 joules. Soldier reaches the state of fatigue when his/her energy lowers to 1500 joules. Table 3
shows the round in which soldier reaches to state of fatigue in all three scenarios.

\begin{table}[!ht]
\begin{center}
\caption{Fatigue of soldier}
  \begin{tabular}{|c|c|c|c|}\hline
   \textbf{Protocol}            & \textbf{Fatigue (round)}       \\ \hline \hline
    \textbf{Walking }           & 10182                           \\ \hline
     \textbf{Slow running}      & 6454                             \\ \hline
      \textbf{Fast running}     & 3811                              \\ \hline
\end{tabular}
\end{center}
\end{table}

\vspace{-1cm}
\section{Conclusion and Future Work}
Energy saving of sensor node is a challenging task for researchers nowadays. In this paper, we proposed an event driven routing protocol for WBASNs. Specifically,
Three senors are attached with soldier's body to monitor his/her heartbeat, body temperature and glucose level in blood. An additional measure is fatigue of the
soldier. Our
 proposed protocol takes three scenarios for measuring the fatigue of soldier i.e. (1) walking, (2) slow running and (3) fast running. Results show that our
  protocol performs better in terms of network lifetime and throughput.

As a part of our ongoing research, we are working on implementing our proposed protocol for more than one soldier. In future,
we are interested to work on MAC layer energy efficient protocol like~\cite{17},~\cite{18},~\cite{19}, ~\cite{20}, ~\cite{21}, etc.

%\bibliography{References}
%\bibliographystyle{IEEEtran}

\end{document}